\documentclass[11pt]{article}
\usepackage{geometry}                

\usepackage{graphicx}
\usepackage{amssymb}
\usepackage{epstopdf}
\newtheorem{unitas}{}[section]
\renewcommand{\theunitas}{\kern-3pt \arabic{unitas}}

\title{
{\large  On 
``law without law"\\
}
}
\author{\normalsize David Ritz Finkelstein\\Georgia Institute of Technology}

\begin{document}
\maketitle

\abstract {
A quantum mechanism for logogenesis is conjectured.}

\section{``Collapse"}

While most physicists use quantum theory in the same way and arrive at the same experimental conclusions, there have been two different theories
since 1924.
Heisenberg's statistical theory of the atom, involving particles-with-non-commuting-properties,
gave the hydrogen spectrum.
Schr\"odinger's wave theory of the atom, involving waves running around the nucleus,  gave the terms of the spectrum
but not the spectrum itself, since it lacked the basic quantum relation $E=\hbar \omega$.
Heisenberg discovered quantum theory while Schr\"odinger discovered  the Schr\"odinger equation.

A vector undergoing the non-commuting operators of the quantum theory
is called a ket or a bra by Dirac, who had doubts about the long-range validity of quantum theory.
His kets and bras represent input beams 
to the experiment and outtake beams from the experiment, respectively,
and the operations that produce them.
Quantum theory uses them as  statistical mechanics uses probability distributions,
so Heisenberg called them probability vectors;
their 
components are actually probability amplitudes.

When quantum systems are combined, their probability vectors combine multiplicatively, much as probability distributions do in statistical mechanics, and not additively, as wave functions of physical wave packets in space do.
Their components are functions of variables of  all the systems, like probability distributions, not just one set of spatial coordinates like wave functions.
Where the predicates of a classical system are represented by subsets of a state space,
those of a quantum system are represented by operators on probability vectors that 
are reproducible ($PP=P$) and symmetric ($P^*=P$).
Like actions in general, they do not commute.
A probability vector defines an irreducible projection operator
($\mbox{Tr } P=1$)  and a homogeneous beam that is put in or taken out in the experiment.

Nevertheless some physicists talk about 
probability vectors as if they described waves as real as the system itself.
Such formulations are called ontological.
The wave ontology helped Schr\"odinger  discover his Equation,
but blocked him from discovering the quantum theory
or ever fully accepting it.
In a later well-known paper he gives a  wave function to a supposedly living 
Cat in a Box
without noting that such a probability vector actually describes 
a  coherent beam of frozen cats near absolute zero.
``Schr\"odinger states" of quantum computers today are cryogenic triumphs; 
a cat in a ``Schr\"odinger state" would be frozen stiff, and the question of life or death that Schr\"odinger wished to ask would  have been answered before the experiment.
The object lesson is that our ordinary intuition deals with highly
disordered systems, while quantum theory 
can cope with extreme order.  
The classical intuition needs retraining, just as for relativity.

Newton too used a wave ontology to
cope with the random behavior of photons at polarizing or reflecting surfaces,
and some teachers of quantum theory today
still do.  

Malus, Heisenberg, Pauli, Schwinger, Feynman, and many others eschewed ontological interpretations and used the statistical interpretation that is now part of the quantum theory.
The electron in a hydrogen atom is 
not a wave but a quantum, a
particle  with non-commuting properties.
Neither an electron interference pattern nor an atomic orbital is an electron, 
any more
than a sound wave in air is an air molecule. 
One probability vector is needed for the input source, another for the outtake
counter, and neither changes during an experiment with one quantum.
By definition, a probability distribution (or vector) is unaffected by what is done to one member of the large population it describes.

A measurement on a system cannot do anything to the probability vectors
describing the system source or sink.

``Wave function collapse" is a non-phenomenon arising from mis-interpretation.

To be sure, quantum field theory has 
waves, whose  phase angles  do not commute with their amplitudes,
and also has quanta, in a complementary relationship,
but this  complementarity is not that between representations of the system as
one wave  or as one particle.

A formulation of quantum theory
in terms of collapsing ``states" that actually exist in the individual atom
was
named by Wigner
the ``orthodox interpretation" and attributed to \cite{NEUMANN1932},
chapter 5, section 1, 
paragraphs (1.) and (2.).
It implies 
\begin{description}
\item [S1] A  single quantum system has a  state,  a ray in its Hilbert space, 
defined by a unit vector $\psi$.
\item [S2] A state $\psi$  determines the probability $\psi^*P\psi$ for every system predicate $P$.
\end{description}
S1 attributes a  state  to a single system.
This is explicit in \cite{NEUMANN1932}, for example in footnote 155.
It is also contradicted in \cite{NEUMANN1932},
when it is pointed out that probability vectors are associated with pure ensembles.
In \cite{NEUMANN1932}
a probability vector is both ontological (S1) and statistical (S2).
This has not proved disastrous in application
because wave theorists are skilled in choosing among inconsistent principles
to get the correct statistical results.
Users of the orthodox formulation   use S1 only in 
discussions, never when applying the theory.
It is never asked how one is to determine the ``state" from a given single quantum.

Heisenberg called his own statistical formulation the 
``Copenhagen Interpretation" by 1955.
It includes S2 but not S1.
In the statistical formulation
one system does not have a probability vector
any more than one system has a probability distribution in statistical mechanics.
Only
a pure  beam of systems,
one that is not a  mixture of statistically distinguishable beams, has one.
There is no physical way to reconstruct a beam of systems from one of its systems,
so the difference is significant.

Therefore
the name ``Copenhagen Interpretation" is now ambiguous.
Let us retain the older terms ``statistical" and ``ontological".

In Newton's experiment and Malus' Law for photon polarization, 
the  beam from the first calcite crystal is 
reduced by the action of the second.
The observer has nothing to do with that process.
``Observer participation" is a by-product of the ontological formulation.

The wave ontology assumes that a complete mathematical model of the system is possible.
Since predicates of symbolic descriptions commute and those of quantum systems do not,
quantum theory forbids such descriptions.
Long before quantum theory, some philosophers (such as Vico) 
noted that it was impossible to say everything about anything in nature.
Probably ontological interpretations survive side-by-side 
with statistical ones
because some physicists still think that the goal of physics
is to describe the universe completely, 
and the probability vector is the only mathematical description at hand.
Others think
that it is to discover what we can do within this universe,
so that we can act wisely.

In his thesis Von Neumann had already converted set theory
from an ontology to a theory of functions,  and so
from objects to non-commutative operations  \cite{NEUMANN1925}.
His functional set theory was founded on Boolean logic, 
but it may have prepared von Neumann  to extract from Heisenberg's operational theory
  the quantum logic  found in  \cite{NEUMANN1932}.

Yet  \cite{NEUMANN1932} is also the source for
the ``orthodox interpretation".
There is  a puzzling double
inconsistency in this; 
one already mentioned within the orthodox formulation itself,
and a second one between that interpretation and the rest of the book.

In conversation, Wigner once mentioned that although
usually the flow of information, as he put it,  was from von Neumann to Wigner,
Wigner
contributed the  cited formulation of the orthodox interpretation
in \cite{NEUMANN1932}.
I gather that the orthodox formulation
was truly Wigner's way of understanding the Heisenberg quantum theory,
and that Wigner believed it to be 
an interregnum theory, 
to be replaced when a physical theory of consciousness was discovered.
The mixed authorship of \cite{NEUMANN1932}
would explain its internal contradictions.
It is consistent with this theory that after 1932 
von Neumann continued to study logics with transition probabilies
while Wigner, for a time, continued to  assign ontological ``state vectors" 
to single systems.

Some students never meet the quantum theory but only the ontological theory,
which some then reject
with good reason.
 To understand quantum theory as Heisenberg did 
demands a greater  language-discipline
than even special relativity,
which still allows complete descriptions,
even for objects outside our light-cone.
Since our best models are intrinsically statistical,
we must renounce complete symbolic models of Nature.
At that moment we leave Descartes and rejoin most of humanity.

\section{Logogenesis} 

Call the process, if any, by which natural laws are formed  ``logogenesis".
Josephson  proposed that quantum observer-participation leads to logogenesis \cite{JOSEPHSON2012}.

There is a well-known example  in pre-quantum physics.
The  geometries of space and time were once generally accepted
as fixed complete  laws of nature, for example by Kant and Poincar\'e. 
Then general relativity provided a theory of 
how these geometric ``laws"
 emerge.
The  ``law"  governing a free-particle trajectory in special relativity
is the geodesic principle in an external gravitational field,
in a singular flat limit that works for any sufficiently  small space-time neighborhood.
This means that the ``law" for any particle is set by the rest of the cosmos.
This phenomenon is governed in its turn, however, by a higher-level law,
Einstein's Equation.

In the logogenesis proposed by Peirce,
nature first acts by chance,
then  acts form habits, and finally habits harden into more permanent laws.
The formation and hardening of habits are 
not further described by Peirce.
I speculate next on a still-unformulated  quantum logogenesis
with elements of those of Einstein and Peirce.
Peirce's ``habit-forming tendency  of  nature"
can be
read as a remarkable premonition of  Bose statistics.

In each step in time,
the system is first annihilated and
then recreated.
This was asserted by Islamic Scholastics of 10th century Baghdad \cite{ABDI1986} and
is  explicit in quantum field theory,
where  a creation $\psi^*$ follows every annihilation $\psi$  in the action principle for a particle.

To create (input) the {\em dual} of a particle is the same act as to annihilate (outtake) the particle.
For example, an anti-particle with positive energy is  the dual to a particle with negative energy.
Any operator that represents one step in time for a particle---for example.
a Hamiltonian operator---is isomorphic in its transformation properties
 to a  probability vector
  for
 a pair of a particle and a dual particle.
Many steps in time mean many such self-dual pairs.
  This statement merely counts indices
  on the operator that defines the dynamical development of the system.
 
Therefore the statistical laws of quantum dynamics, 
 from Heisenberg's equations of motion to the action principles of 
 Dirac, Feynman, and Schwinger,
have the same mathematical form as that specifying
the transition probability amplitude
between one  probability vector for the
experimental process
and one for the dynamics, the ``law" of motion.

For example, a transition probability amplitude 
$A=<2|H|1>$ is also the transition probability amplitude between the
experimental pair   $|1><2|$  and the dynamics  pair $H$.

But in quantum theories,  probability vectors  generally represent  beams or sources.
Where in the world is the beam that a dynamics probability vector like $H$ could represent?
    
The source of any quantum system  is  the co-system, the rest of the universe.
Therefore
the dynamics probability vector $H$ might merely describe the co-system, in just enough detail for the experiment under study.
What governs  any system would then be the rest of the universe,
in quantum theory as well as in gravity theory.
The law is the other.

This relies, first, on the fact that in quantum experiments
of maximal resolution
the system is a minuscule part of the universe.
This leaves room for its law  in the co-system.
If the quantum system could be the entire universe, 
its observer and its  law would have to ``come from outside the universe".

The rest of the universe is mostly vacuum, with many virtual pairs.
To be described by a pure one-system probability vector,
 it must be a coherent Bose condensate,
organized like the BCS model of a superconductor.
That is also why our blunt macroscopic controls can
produce sharp beams of quantum systems.

It is then permissible to ask
what organizes the rest of the universe so coherently.
Is there a higher-level dynamics, analogous to Einstein's equations?
If we ask this question experimentally, however, the co-system,
or at least the minuscule part of it that we can observe sharply,
becomes the system under study.
Then we have already explained its organization.
Each small part of the universe is thus influenced by
the rest through quantum statistics; that is  the conjecture.
This might be expressible as a self-consistency
condition on a history probability vector.

Then  Nature would be 
ruled by  chance, Peirce's Tyche,
each decision being made by an individual elected by lot for the occasion;
but now quantum chance, not classical.
Such a ``tycheocracy" could run smoothly 
if it is mainly a supercondensate of
quantum pairs that
cannot be told apart by their actions.

There would likely be  disorganized regions above the critical temperature,
 where  space-time and law melt down.
Such weak links might even be Josephson junctions in a generalized sense.
 One naturally conjectures them at  the cores of Big Bangs and black holes.

\section{Peircean semiotics}

Peirce's signs occur in a semantic triangle
of sign, interpretant, and object. 
For the ardent evolutionist Peirce, moreover,
what makes a triangle  semantic is
just its tendency to reproduce and survive natural selection,
as when babies learn their parents' languages.
Peirce puts physics into biology more than the converse.
In another example among many,  
a DNA molecule is the sign,
an RNA polymerase is the interpretant, the resulting RNA molecule is the object,
and  the cell is the ``immediate system of interpretance" 
\cite{SALTHE2009}.
As this example shows, the present quantum theory
can still be expected to cover semiotic processes satisfactorily if
 the openness and organization of the systems in commmunication
 are duly recognized.

 Peirce's pragmatism 
seems to be appropriate  for
the study of quanta, as has been  pointed out 
\cite{STAPP1972}.
On the other hand,  he ignored both relativity and quantum theories, as far as I can find,
and believed in apriori truths that they invalidate.
His First Flash creates  Matter 
but not Space and Time, which  he took to be forever flat.
His synechism expressed his deep belief
  in real continua and infinities, a belief which hardly seems pragmatic.
As far as I know, he did not explicitly propose, as Boole had proposed early on,
that even Boole's laws of thought might break down in practice.

Moreover, Peirce's theories of evolution seems to perpetuate a serious omission of Darwin.
The small random changes considered by Darwin must be supplemented by large organizing changes, the symbiogenesis 
observed by Lynn Margulis \cite{MARGULIS2002} in biological evolution and the modular architecture observed by
Herbert Simon in
computation.  Modular architecture is crucial in language too.

Any physicist starting over today from the ideas of Peirce
would have to build a road to the well-tested ideas of Einstein, Heisenberg,
and the Standard Modelers.
It seems more pragmatic to go onward from them than back to Peirce.
We may still
draw on his remarkable spirit when we need it.

 I thank Heinrich N. Saller and Stanley N. Salthe for useful comments.

\bibliography{DRFbib20110805}
\bibliographystyle{plain}

\end{document}